\documentclass[%
 preprint,
 amsmath,amssymb,
 aps,
]{revtex4-2}

\usepackage{graphicx}
\usepackage{dcolumn}
\usepackage{bm}
\usepackage{color}

\begin{document}
\preprint{APS/123-QED}

\title{Statistical Representation of Spacetime}

\author{Hamidreza Simchi}
\email{simchi@alumni.iust.ac.ir}
\affiliation {Department of Physics, Iran University of Science and Technology, Narmak, Tehran 16844, Iran} \affiliation{Semiconductor Technoloy Center, P.O.Box 19575-199, Tehran, Iran}

\date{\today}
\begin{abstract}
It is assumed that the spacetime is composed by events and can be
explained by partially ordered set (causal set). The parent events born
two kinds of children. Some children have a causal relation with their
parents and other kinds have not. It is assumed that evolution of the
population is only happen by the causal children. The assumed population
can be modeled by finite (infinite) dimension Leslie matrix. In both
finite and infinite cases, it is shown that the stationary state of the
population always exists and the matrix has positive eigenvalues. By
finding the relation between the statistical information of the
population and the stationary state, a probability matrix and a
Shannon-like entropy is defined. It is shown that the change in entropy
is always quantized and positive and in consequence, the world is
inflating. We show that the vacuum energy can be attributed to the
necessary done work for preserving the causal relation between the
parents and the children (cohesive energy). By assuming that the sum of
cohesive energy and kinetic energy of the denumerable causal spacetime
is equal to the heat, which flows across a causal horizon, we find the
relation between energy-momentum tensor and discrete Ricci tensor which
can be called the Einstein state equation. Finally, it is shown that the
constant of proportionality \(\eta\) between the entropy and the area is
proportional to \(\frac{k_{B}}{l_{p}^{2}}\) at Planck scale which is in
good agreement with the Hawking's result.
\end{abstract}

\maketitle


\section{Introduction}
After the discovery of quantum physics, it became possible to describe
the electronic and structural properties of materials based on their
constituent elements, namely atoms and electrons. Using the concepts
related to atomic orbitals and their hybridization, the cause of the
stability of the material structure was described {[}1{]}, and based on
the energy band theory, and many-body theory, the electronic properties
and the quantum transport properties of the majority carriers were well
described {[}2-5{]} so that the computational results were consistent
with what was seen in the laboratory. This process continued with the
discovery of more elementary particles, and scientists tried to describe
strong nuclear forces {[}6{]} and weak forces {[}7{]} based on the
properties of related elementary particles. Hence, it can be said that
quantum physics, many-body theory, and quantum field theory are attempts
to express and interpret the macroscopic properties of a physical system
based on its atomic and sub-atomic microscopic components. Of course, it
should be noted that many efforts were made to unifying the known forces
in nature so that finally, by presenting a standard model {[}8{]}, these
efforts were completed, without considering the force of gravity.

In contrast, the theory of relativity is not an attempt to describe the
macroscopic properties of a system based on its microscopic properties.
This theory can be considered as a description of how energy and matter
interact with the gravitational field (spacetime field) {[}9-10{]}. It
is based on how spacetime field interacts with matter that the planets
move in a certain orbit, light is deflected when it passes near the sun,
the existence of black holes in the galaxy is predicted, and a
scientific description is given for some other cosmic phenomena. Of
course, the efforts made to determine the cause of black hole radiation
{[}11{]} and the thermodynamic description of Einstein's relativistic
equation {[}12{]}, which is made by using a combination of quantum
physics and Einstein's theory of relativity, must be taken into account.

If we take quantum gravity theory as a theory for describing the
observable macroscopic phenomena based on the atomic and sub-atomic
microscopic components that make up the spacetime field, then when the
problem scale approaches to the Planck scale, the question arises as to
what these microscopic components are and what are their
characteristics? It is well known that one of the important features of
worldline is its causal property. This means that at any point of
spacetime (present) one set of events can be considered as past and
another set as future event related to the present. Due to the constant
and maximum speed of light in the vacuum and this causal relationship
between the present, the future and the past, some events will not have
any physical connection with the present. In order to mathematically
represent this causal property of events, the partially ordered causal
set, which has been studied by mathematicians before {[}13{]}, may be
used.

Attempts to use partially ordered sets as the microscopic elements that
make up spacetime go back many years. By defining the null, parallel
lines and planes and proving numerous theorems involving them, Robb has
described the relativity using the discrete spacetime (i.e., casual
structure) {[}14,15{]}. It has been shown that the casual structure of a
spacetime, together with a conformal factor, determine the metric of a
Lorentzian spacetime, uniquely {[}16,17{]}. Therefore, one can recover
the conformal metric by using the before and after relations amongst all
events {[}18{]}. Now, if one has a measure for the conformal factor,
he/she can recover the entire metric and spacetime {[}18{]}. Of course,
\textsuperscript{,}t Hooft {[}19{]} and Myrheim {[}20{]} have
independently found the causal set theory too. Extensive research is
being done today on the theory of causal sets and its application in the
formulation of the theory of relativity at the Planck scale {[}21{]}. In
one group of this research, by focusing on the causal events, an attempt
is made to study the static and dynamic state characteristics of the
causal set {[}22, 23{]}, while a bunch of research focuses on the causal
relationship between events {[}24{]}. Of course, the third group, by
assuming the irreversibility of time and explaining and using the
principles of energy and momentum conservation and the absence of red
shift in the momentum relation at Planck scale, have introduced
energetic causal set theory that includes both classical and quantum
cases {[}25, 26{]}. Ref. 21 is a fine review on the causal set theory
and its outlook.

Now, the question that can be asked is whether spacetime can be
considered as a population consisting of discrete members (parents)
between whom there is a causal relationship (such as a partially ordered
set) that has evolved with the birth of new members (children) of this
set and \(i\)-th step goes to \((i + 1)\)-th step and define entropy for
this set and study it, statistically? In this article, we try to answer
this question in the affirmative. We assume that spacetime is composed
by events which is shown by the partially ordered set. At each \(i\)-th
step, the ordered set is called the parent set, and at each
\((i + 1)\)-th step, the ordered set is called children set. Some
children have causal relation with their parents and some have not. It
is assumed that the evolution of the system is only done by causal
children. By using finite and infinite Leslie matrix, the system is
modeled and it is shown that the stationary state always exists. We show
that, since for preserving the causality relation between parents and
children some works should be done, the entropy of the system is
quantized and increases by the birth of new children. It means that we
encounter the universe inflation. Finally, by using the Jacobson's
procedure {[}12{]}, we calculate the energy-momentum tensor and in
consequence the Einstein state equation including discrete Ricci tensor.

The structure of the article is as follows. The basic assumptions are
introduced in section II and the Leslie matrix model is introduced in
section III. The quantized entropy and Einstein state equation are
studied in section IV and V, respectively. The summary is provided in
section VI.

\section{Assumptions}

Let us to consider a causal set \(C\) which is a locally finite,
partially ordered set. The order relation is shown by `\(\prec\) ` and
the set has the below properties {[}13, 22{]}:

\begin{enumerate}
\def\labelenumi{\arabic{enumi}.}
\item
  For all \(x \in C\) , \(x \nless x\) (Irreflexive)
\item
  For all \(x,y,z \in C\) if \(x \prec y\) and \(y \prec z\) then
  \(x \prec z\) (Transitive)
\item
  For all \(x,y \in C\) if \(x \prec y\) and \(y \prec x\) then
  \(x = y\) (Antisymmetric)
\item
  For all \(x,y \in C\) we have
  \(\left| \left( z \in C \middle| x \prec z \prec y \right) \right| < N_{0}\)
  (Locally finite)
\end{enumerate}

The causal set can be considered as age-structured populations if adding
elements to the causal set are considered as individuals born. One of our main goals is mixing the Sorkin’s idea [23] about the new generations in the causal set theory and the Demetrius’s idea [30] about the aged-structure populations model for developing own model. Demetrius has begun with the Leslie model and considered a population divided into $n$ age classes [30]. Sorkin has assumed that the causal sets can be formed by adjoining a single maximal element to a given causal set and the result will be called a family [23].For doing that, we used the rules which are mentioned in table one. 

{\begin{table}[ht]
\caption{Necessary rules for developing the new model} 
\centering 
\begin{tabular}{c c c c} 
\hline\hline 
Rule & Sorkin’s(S) Or Demetrius’s(D) idea & New status (our model) & Figure \\ [1ex] 
\hline 
$1$ & Timid children(S) & Causal children & Red cycles in Fig.1 \\ 
$2$ & Gregarious child(S) & Non-causal children & Blue cycles in Fig.1  \\
$3$ & $m_i$-individuals born numbers(D) & $m_i$-causal children numbers & Green box in Fig.2 \\
$4$ & $b_i$-proportion of surviving individuals(D) & $b_i$-proportion of non-causal children & Red box in Fig.2 \\
$5$ & All children are next parents(S) & Causal children are next parents & Fig.1\\
\hline 
\end{tabular}
\label{table:nonlin} 
\end{table}

 At each step of growth, two kinds of babies may be born. In the first kind,
\(m_{i}\) number of individuals born have the order relation with their
parents and in the second kind, some individuals born have not. For the
second kind, \(b_{i}\) stands for the proportion of individuals born.
The individuals born with order relation are placed on the causal
worldline and the individuals born without order relation are not
{[}27{]}. It is assumed that the evolution of the population is only
done by causal children. As Fig.1 shows, one can find the number of
causal individuals born (non-causal individual born) by counting the red
(blue) cycles from the beginning up to the \(i\)-th step. All of the red
cycles may be considered as parents in the \(i\)-th step.

Based on the mentioned rules in table 1, the number of parents increase by increasing the growth steps.  In fact, the new causal children in $i$-th step are considered as new parents of the children in $(i+1)$ -th step. Therefore, the change in age structure between step $i$ and step $(i+1)$  can be found by using the Leslie matrix as Demetrius has assumed [30]. It means that, $m_1$ stands for the number of causal children in step one respect to itself, $m_2$ stands for the number of causal children in step two respect to step one and generally, $m_i$ stands for the number of causal children in the $i$-th step respect to step one. Also, $b_1$ stands for the proportion of non-causal children in step one surviving to step two, $b_2$ stands for the proportion of non-causal children in step two surviving to step three, and generally $b_j$ stands for the proportion of non-causal children in step $j$ surviving to step $(j+1)$.

\begin{figure}[]
\includegraphics[width=\columnwidth]{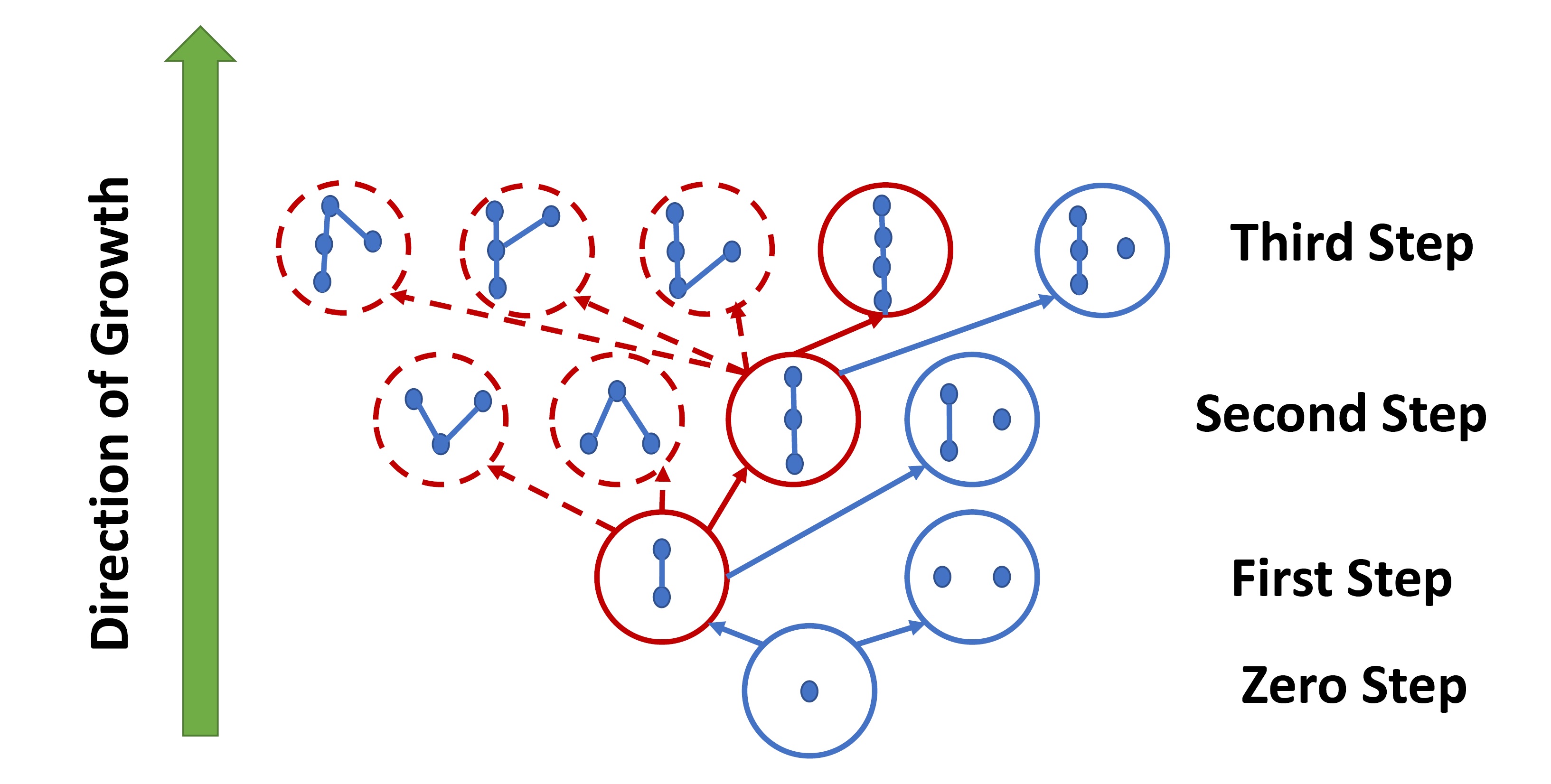}
\caption{\label{fig:epsart}  \textbf{(Color online) An example of a causal set as a population
growth. The events in red cycles are placed on the causal worldlines.
The non-dashed red color cycles show an example of the casual worldline.
The blue cycles show the non-causal events. One can find the number of
causal individuals born (non-causal individuals born) by counting the
red (blue) cycles from the beginning up to the i-th step.}}
\end{figure}

\section{Matrix Representation}

\noindent
 Let us to define the below Leslie matrix {[}28{]}
\begin{equation}
M_{\text{ij}} = \left\{ \begin{matrix}
m_{j}\  > 0\ \ \ \ \ \ for\ i = 1 \\
0 < b_{i} \leq 1\ \ \ \ \ for\ i = j + 1 \\
0\ \ \ \ \ \ \ \ \ \ \ \ \ \ \ \ \ Otherwise \\
\end{matrix} \right.\  
\end{equation}
i.e.,
\begin{equation}
M = \begin{pmatrix}
\begin{matrix}
m_{1} & m_{2} & m_{3} \\
b_{1} & 0 & 0 \\
0 & b_{2} & 0 \\
\end{matrix} & \begin{matrix}
m_{4} & m_{5} & \cdots \\
0 & 0 & \cdots \\
0 & 0 & \cdots \\
\end{matrix} & \begin{matrix}
\cdots & \cdots & m_{n} \\
\cdots & \cdots & \cdots \\
\cdots & \cdots & \cdots \\
\end{matrix} \\
\begin{matrix}
0\ \ \  & 0 & \text{\ \ \ \ }b_{3} \\
0 & 0 & \ \ 0 \\
0 & 0 & \ \ 0 \\
\end{matrix} & \begin{matrix}
0 & 0 & \cdots \\
b_{4} & 0 & \cdots \\
0 & b_{5} & \cdots \\
\end{matrix} & \begin{matrix}
\cdots & \cdots & \cdots \\
\cdots & \cdots & \cdots \\
\cdots & \cdots & \cdots \\
\end{matrix} \\
\begin{matrix}
0\ \  & 0 & \ \ \ 0 \\
 \vdots & \vdots & \vdots \\
 \vdots & \vdots & \vdots \\
\end{matrix} & \begin{matrix}
0 & 0 & b_{6} \\
 \vdots & \vdots & \vdots \\
 \vdots & \vdots & \vdots \\
\end{matrix} & \begin{matrix}
\cdots & \cdots & \cdots \\
 \vdots & \vdots & \vdots \\
 \vdots & b_{n - 1} & 0 \\
\end{matrix} \\
\end{pmatrix}
\end{equation} 
\\
\\
\\
\\
Generally, based on the Perron--Frobenius theorem {[}29, 30{]}, The
eigenvalues of \(M\) are real and it has eigenvalues with all
positive elements such that
$M\vec{u}=\lambda\vec{u}$ and $\vec{v}M=\lambda\vec{v}$.
\\
\\
It should be  noted that, one of the important subjects is the relation between the elements of the Leslie matrix and the future growth of the population {[}28, 30{]}. It has been shown that there are relationships between certain elements of a population and the dominant eigenvalue, which determines growth {[}28, 30{]}.  For example, If the dominant eigenvalue and, hence, all the eigenvalues are less than 1, then the population will decline. If the dominant eigenvalue is greater than one, regardless of the values of the other eigenvalues, the population will grow {[}28, 30{]}. In below, the population growth will be related to the world inflation. } 
It has been shown that, if {[}30{]}
\\
\begin{equation}
l_{j} = \left\{ \begin{matrix}
1\ \ \ \ \ \ \ \ \ \ \ \ \ \ for\ \ \ \ \ \ \ j = 1 \\
\Pi_{r = 1}^{j - 1}\text{\ \ }b_{r}\ \ \ \ \ \ for\ j \geq 2 \\
\end{matrix} \right.\ 
\end{equation}
\\
Then
\begin{equation}
u_{i} = \frac{l_{i}}{\lambda^{i}}\ \ \ \ ,\ \ \ v_{i} = \frac{\left( \sum_{j = i}^{n}{m_{j}u_{j}} \right)}{u_{i}}
\end{equation}
For example, if
\(M=  \begin{bmatrix}
    m_{1} & m_{2} \\
    b_{1} & 0
  \end{bmatrix}
\)
then \(u_{1} = 1/\lambda_{i}\), \(u_{2} = b_{1}/\lambda_{i}^{2}\),
\(v_{1} = \lambda_{i}\) and \(v_{2} = m_{2}\) (Appendix A).
\\
\\
Now, a question can be asked: what can be the physical interpretation of
\(M,\) \(\vec{u}\) , and \(\lambda_{i}\)?
\\
\\
Fig.2 shows the new causal and non-causal children of \((i+1)\)-th generation step when
the \(n_i\) causal parents present at \(i\)-th generation step. The parents belong
to the own \(i\)-th step and the all of the previous steps from the
beginning. By substituting Eq. (4) in the eigenvalue equation, one can find
\\
\\
\begin{widetext}
\begin{equation}
\begin{pmatrix}
m_{1} & m_{2} & \begin{matrix}
\cdots & \text{\ \ }m_{n} \\
\end{matrix} \\
b_{1} & 0 & \begin{matrix}
0\  & \ \ 0 \\
\end{matrix} \\
\begin{matrix}
0 \\
0 \\
\end{matrix} & \begin{matrix}
\cdots \\
\cdots \\
\end{matrix} & \begin{matrix}
\begin{matrix}
0\ \ \  & 0 \\
\end{matrix} \\
\begin{matrix}
b_{n - 1} & 0 \\
\end{matrix} \\
\end{matrix} \\
\end{pmatrix}\begin{pmatrix}
1/\lambda \\
b_{1}/\lambda^{2} \\
\begin{matrix}
 \vdots \\
b_{1}\cdots b_{n - 1}/\lambda^{n} \\
\end{matrix} \\
\end{pmatrix} 
= \lambda\begin{pmatrix}
1/\lambda \\
b_{1}/\lambda^{2} \\
\begin{matrix}
 \vdots \\
b_{1}\cdots b_{n - 1}/\lambda^{n} \\
\end{matrix} \\
\end{pmatrix}
\end{equation}
\end{widetext}
or
\begin{widetext}
\begin{equation}
\begin{pmatrix}
\frac{m_{1}}{\lambda} + \frac{m_{2}b_{1}}{\lambda^{2}} + \cdots + \frac{m_{n}b_{1}b_{2}\cdots b_{n - 1}}{\lambda^{n}} \\
b_{1}/\lambda \\
\begin{matrix}
 \vdots \\
b_{1}\cdots b_{n - 1}/\lambda^{n - 1} \\
\end{matrix} \\
\end{pmatrix} = \begin{pmatrix}
1 \\
b_{1}/\lambda \\
\begin{matrix}
 \vdots \\
b_{1}\cdots b_{n - 1}/\lambda^{n - 1} \\
\end{matrix} \\
\end{pmatrix}
\end{equation}
\end{widetext}
Therefore, he/she can write the below eigenvalue equation
\begin{equation}
\begin{pmatrix}
\frac{m_{1}}{\lambda} & \frac{m_{2}b_{1}}{\lambda^{2}} & \text{\ \ \ }\begin{matrix}
\cdots & \frac{m_{n}b_{1}b_{2}\cdots b_{n - 1}}{\lambda^{n}} \\
\end{matrix} \\
1 & 0 & \begin{matrix}
\cdots\ \ \ \ \ \ \ \ \ \ \ \ \ \ \ \  & 0 \\
\end{matrix} \\
\begin{matrix}
0 \\
0 \\
\end{matrix} & \begin{matrix}
\cdots \\
\cdots \\
\end{matrix} & \begin{matrix}
\begin{matrix}
\cdots \\
1 \\
\end{matrix}\text{\ \ \ \ \ \ \ \ \ \ \ \ \ \ \ \ } & \begin{matrix}
0 \\
0 \\
\end{matrix} \\
\end{matrix} \\
\end{pmatrix}\begin{pmatrix}
1 \\
1 \\
\begin{matrix}
 \vdots \\
1 \\
\end{matrix} \\
\end{pmatrix} = \begin{pmatrix}
1 \\
1 \\
\begin{matrix}
 \vdots \\
1 \\
\end{matrix} \\
\end{pmatrix}
\end{equation}

It means that he/she can define the probability element and the
probability matrix as below {[}30{]}
 \\
 \begin{equation}
p_{i} = m_{i}l_{i}/\lambda^{i}
\end{equation}
\begin{equation}
P=(P_{\text{ij}}) = \left\{ \begin{matrix}
p_{i}\ \ for\ i = 1 \\
1\ \ \ for\ i = J = 1 \\
0\ \ \ \ \ \ \ \ \ \ Otherwise \\
\end{matrix} \right.\ 
\end{equation}

\begin{figure}[]
\includegraphics[width=\columnwidth]{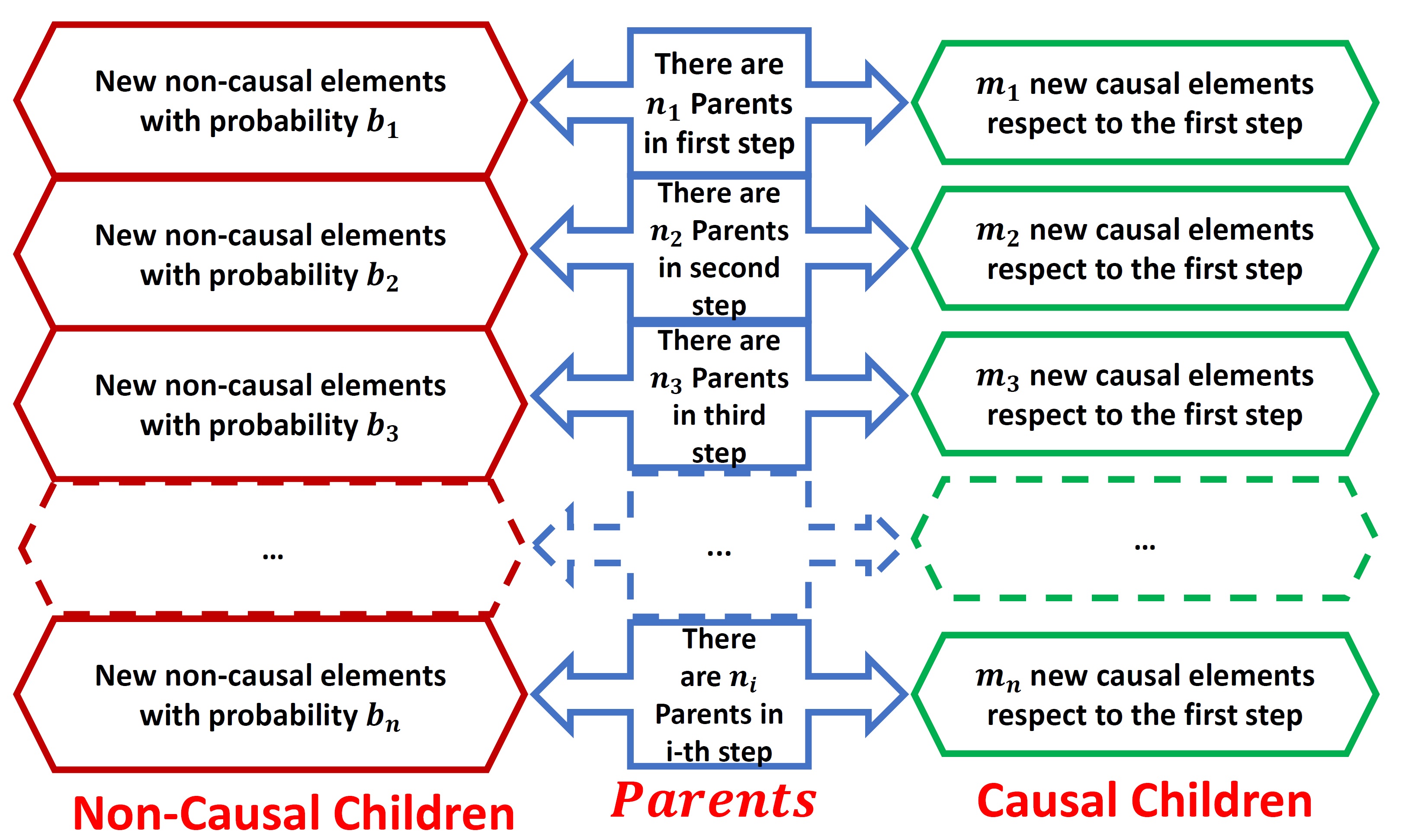}
\caption{\label{fig:epsart}  \textbf{(Color online) The relation between
i-th and (i+1)-th step.}}
\end{figure}
 
Therefore, \(m_{i}u_{i} = m_{i}l_{i}/\lambda^{i}\) can be considered as
the born probability of \(m_{i}\) new causal children related to the
parent \(u_{i}\) and in consequence
\(m_{1}u_{1} + m_{2}u_{2} + \cdots + m_{n}u_{n}\) is the total born
probability of new casual children at (\(i\) +1)-step.

By attention to the above descriptions, the matrix \(M\) is the
evolution matrix such that
$\vec{u}^{\prime}(i+1)=M\vec{u}(i)$
where
$\vec{u}(\vec{u}^{\prime})$
is the
distribution of parents from the beginning up to the \(i\) (\(i\) +1)-th
step. It should be noted that if the term
\(M_{\text{nn}} = b_{n} \neq 0\) it means that the n-th parent born
non-causal child with the probability \(b_{n}\) {[}28{]}. The obtained
general results satisfy again and one should only find the new
eigenvalues and eigenvectors. For providing the physical interpretation
of \(\lambda\) we define the entropy in the next section.
   
\section{Entropy}

\noindent

Entropy is the total number of ways to rearrange the internal
microstates of a system while keeping its external macrostates unaltered
{[}31{]}. Here, we encounter the ensemble of parents with the boron
probability \(p_{\text{ij}}\) as internal microstates and the stationary
population as the macrostate. For studying the effect of the variation
of microstates on the macrostate, we should define the entropy.

By using the irreduciblity properties of matrix $P$ and associating the graph $G(P)$ to the matrix $P$, Demetrius {[}30{]} has represented the set of all paths in the graph by $\Omega$. He also defined the shift transformation $T$ as  $\Omega$ $\rightarrow$ $\Omega$  and called $(\Omega,T)$ the symbolic dynamical system associated with the matrix $P$. He has used the Markov matrix $P$ to introduce a probability measure $\mu$ on the space of sequences $\Omega$ and finally defined the entropy $H(T)$$[$30$]$.

By using the Eq. (4) we can define the new population vector
$\vec{Z}$ as below {[}30{]}
 \\
 \begin{equation}
Z_{i} = \frac{1}{\sum_{i = 1}^{n}{ip_{i}}}u_{i}v_{i} = \frac{\sum_{i = 1}^{n}p_{i}}{\sum_{i = 1}^{n}{ip_{i}}} = \frac{1}{\sum_{i = 1}^{n}{ip_{i}}}
\end{equation}
 \\
Therefore, $\vec{Z}P = \vec{Z}$. Then,
$\vec{Z}$ stands for the stationary distribution and in
consequence, similar to the Demetrius's method, we define the population entropy for the stationary
distribution as below {[}30{]}
\\
\begin{equation}
S = - \frac{\sum_{i = 1}^{n}{p_{i}\ log(p_{i})}}{\sum_{i = 1}^{n}{ip_{i}}}
\end{equation}
\\
where, \(p_{i} = m_{i}l_{i}/\lambda^{i}\). It should be noted that the
numerator of the Eq. (11) is similar to the Shannon entropy {[}32{]}.
For providing the physical interpretation of \(\lambda\) we write
\\
\begin{widetext}
\begin{equation}
log\lambda = \frac{\sum_{i = 1}^{n}{p_{i}log(\frac{\lambda^{i}p_{i}}{p_{i}})}}{\sum_{i = 1}^{n}{ip_{i}}} = \frac{1}{\sum_{i = 1}^{n}{ip_{i}}}\sum_{i = 1}^{n}{p_{i}(log\lambda^{i}p_{i} - logp_{i})}
\end{equation}
\end{widetext}
or
\\
\begin{equation}
log\lambda = \frac{\sum_{i = 1}^{n}{p_{i}\log\lambda^{i}p_{i}}}{\sum_{i = 1}^{n}{ip_{i}}} - \frac{\sum_{i = 1}^{n}{p_{i}\log\left( p_{i} \right)}}{\sum_{i = 1}^{n}{ip_{i}}} = \Phi + S
\end{equation}
or
\begin{equation}
\delta = \Phi + S = \Phi + \frac{S^{'}}{T} = \Phi + T^{'}S^{'}
\end{equation}
where
\(\Phi = \frac{\sum_{i = 1}^{n}{p_{i}\log\lambda^{i}p_{i}}}{\sum_{i = 1}^{n}{ip_{i}}}\),
\(S^{'} = \sum_{i = 1}^{n}{p_{i}\log\left( p_{i} \right)}\),
\(T^{'} = \left( T = \sum_{i = 1}^{n}{ip_{i}} \right)^{- 1}\), and
\(\delta = log\lambda\). From thermodynamic, we know, \(F = U - TS\)
where \(F\) and \(U\) are Helmholtz free energy and internal energy,
respectively. Also, \(T\) and \(S\) are absolute temperature and
entropy, respectively. Since, \(\delta = \Phi + T^{'}S^{'}\) then
{[}30{]}
\begin{equation}
0 = \mathrm{\Delta}\Phi + T^{'}\mathrm{\Delta}S^{'}
\end{equation}
or
\begin{equation}
\frac{1}{T^{'}} = T = \frac{\mathrm{\Delta}S^{'}}{- \mathrm{\Delta}\Phi} \rightarrow \mathrm{\Delta}S^{'} = - T\mathrm{\Delta}\Phi
\end{equation}
In thermodynamic
\(\frac{1}{T} = \frac{\mathrm{\Delta}S}{\mathrm{\Delta}U}\) and in
consequence the temperature of any object can be described by the amount
of heat that must be added to it to increase its entropy by one unit.
Similarly, by attention to the Eq. (16), \(T\) can be described by the
amount of \(- \mathrm{\Delta}\Phi\) that must be added to the population
to increase its entropy by one unit. By comparison between Eq. (14) and
the equation \(F = U - TS\), one can conclude the comparison table 1.

\begin{table}[ht]
\caption{Comparsion between statistical representation and thermodynamic} 
\centering 
\begin{tabular}{c c} 
\hline\hline 
Statiatical Representation & Thermodynamic \\ [0.5ex] 
\hline 
$\delta$ & Helmholtz Free energy \\ 
$T$ & Inverse of absolute temperature \\
$-T^{'}S^{'}$ & Entropy \\
$\Delta\Phi$ & Internal (mean) energy \\
\hline 
\end{tabular}
\label{table:nonlin} 
\end{table}

Now, let us to assume that the causal relation between children and
parents is energetic relation. That is, after establishing a causal
relationship between a parent and a child if we assume that the amount
of \(- \mathrm{\Delta}\Phi\) increases by \(V_{i}\) then
\(\mathrm{\Delta}S = \left( \sum_{i = 1}^{n}{ip_{i}} \right)^{- 1}\sum_{i}^{}{V_{i} > 0}\).
Therefore, the entropy increases by growing the population and in
consequence the population will be more stable than before. Since, the
causal events are denumerable then \(V_{i}\) is discrete and in
consequence \(\mathrm{\Delta}S\) is too i.e., \(\mathrm{\Delta}S\) is
quantized. It can be assumed that, the cohesive energy of the
denumerable causal spacetime is the energy of vacuum which is appeared
in quantum filed theory as \(\sum_{i}^{}{\hslash\omega/2}\). In
consequence, if it is assumed that \(V_{i} = \hslash\omega/2\), since
\(\sum_{i}^{}{V_{i} =}m\hslash\omega/2\). It means that the entropy is
quantized and equal to
\begin{equation}
S = + \frac{m\hbar}{2}T= + \frac{m\hbar}{2}\left( \sum_{i = 1}^{n}{ip_{i}} \right)^{- 1}
\end{equation}
 \\
 where, \(\left( T = \sum_{i = 1}^{n}{ip_{i}} \right)^{- 1} = cte\). For
comparison, we can take into account the formation of a crystal. By
bringing an element of infinity closer to one element in the crystal,
the two elements are bounded to each other and the internal (cohesive)
energy will be increased and stored in the crystal {[}33{]}. It means
that the cohesive energy is always negative {[}33{]}. Similarly, by
adding a new child to the population and establishing the causal
relation the amount of \(- \mathrm{\Delta}\Phi\) increases by the quanta
of energy and stored in the population. Therefore, it is expected that
the energy
\(- \mathrm{\Delta}\Phi = \sum_{i}^{}{V_{i}( =}m\hslash\omega/2)\) will
be equal to the energy of vacuum. Of course, the causal spacetime is a
disordered system and Bloch's theorem does not satisfy here. Increment
of entropy, due to the increasing the number of parents and causal
children, cases the expansion of the universe (universe inflation). Why
we see universe inflation because its entropy increases due to the
increment of the spacetime events.
\section{Einstein State equation}

\noindent

It has been shown that, when \(M\) is an infinite-dimensional population
matrix, under special conditions, a stationary distribution exists
{[}34{]} (Appendix B). Also, when, the index \(i,j \rightarrow \infty\),
we encounter an infinite denumerable Leslie matrix, \(M\). It has been
shown that for \(i,j \rightarrow \infty\), an essentially unique
stationary distribution exists {[}35{]} and there is a generating
function \(\mathbf{G}(z)\) for calculating \(M^{n}\), where \(z\) is a
complex indeterminate. Now, if \(u_{0}\) is the first parent (initial
condition) then \(u_{n} = M^{n}\ u_{0}\){[}35{]} (Appendix B).
Therefore, the stationary distribution exists when
\(i,j \rightarrow \infty\) {[}34, 35{]}.

Also, in spacetime dynamic, it can be assumed that the heat is energy
which flows across a causal horizon and the heat flux is given by
{[}12{]}
\begin{equation}
\delta Q = \int_{}^{}{T_{\text{ab}}\chi^{a}d\Sigma^{b}}
\end{equation}
where, all integrands are defined in Appendix C. It should be noted that
the past horizon of a local Rindler horizon (system) is instantaneously
stationary (in local equilibrium) at spacetime point \(p\) {[}12{]}. We
assume that each unit area of spacetime is composed by the denumerable
causal events. Each causal event has kinetic energy \(K_{i}\) and is
bounded to another causal event by the causal correspondence energy
\(V_{i}\). It can be assumed that the heat which flows across a causal
horizon is equal to
$\sum_{i}^{}\left( V_{i} + K_{i} \right)\delta{A}$ 
where $\delta{A}$ is defined in Appendix C. The assumed phenomenon is similar to the relaxed
state of a crystal. In the crystal, the elements vibrate around their
equilibrium with preserving the bounding between elements (preserving
the cohesive energy). Here, although the unit area of the causal
spacetime is a disorder system and the Bloch's theorem does not satisfy
we assume that
\begin{equation}
\delta Q = - \sum_{i}^{}\left( V_{i} + K_{i} \right)\delta{A}
\end{equation}
Since, the area $\delta{A}$ is composed by the denumerable causal
events by using Eq. (C2) and Eq. (19), we can write
\begin{equation}
- \kappa\int_{}^{}\lambda\ R_{\text{ab}}k^{a}k^{b}{d\lambda dA} = \{ - \sum_{i}^{}{\left( V_{i} + K_{i} \right)\}}\delta{A}
\end{equation}
where, all integrands, $\delta A$ and $\kappa$ are defined in
Appendix C.

Using Eq. (C5)
\begin{equation}
\kappa\int_{}^{}{\text{\ T}_{\text{ab}}k^{a}k^{b}{d\lambda dA}} = - \{\sum_{i}^{}{\left( V_{i} + K_{i} \right)\}}\int_{}^{}{R_{\text{ab}}k^{a}k^{b}{d\lambda dA}}
\end{equation}
Therefore,
\begin{equation}
\kappa T_{\text{ab}}k^{a}k^{b} =
- \{\sum_{i}^{}{\left( V_{i} + K_{i} \right)\}}
R_{\text{ab}}k^{a}k^{b}
\end{equation}
which is valid for all null \(k^{a}\). It implies that {[}12{]}
\begin{equation}
T_{\text{ab}} = - \frac{1}{\kappa}\sum_{i}^{}\left( V_{i} + K_{i} \right)\{ R_{\text{ab}} + \left( - \frac{R}{2} + \Lambda \right)\} g_{\text{ab}}
\end{equation}
 where, \(R_{\text{ab}}\) and \(R\) are Ricci tensor and Ricci scalar,
respectively and \(\Lambda\) is some constant. Therefore, we can imagine
the unit area is composed by the energetic spacetime events with
discrete geometry. As Eq. (23) shows, the energy-momentum tensor can be
calculated by the product of the energy stored within the discrete
geometry and the discrete curvature of the unit area. It means that
energy-momentum and the stationary state as two macroscopic parameters
can be studied by the behavior of the energetic spacetime events which
are the microscopic constitutes. Eq. (23) can be called Einstein state
equation. Th Eq. (23) can be written in matrix form as follows
\begin{equation}
T = Trace\left( \Xi \right)\left( R + \Upsilon g \right)
\end{equation}
 where, \(\Upsilon = - \frac{R}{2} + \Lambda\) and
\begin{equation}
\Xi = - \frac{1}{\kappa}\begin{pmatrix}
V_{1} + K_{1} & 0 & \cdots \\
0 & V_{2} + K_{2} & 0 \\
 \vdots & \vdots & \ddots \\
\end{pmatrix}
\end{equation}
Each \(V_{i} + K_{i}\) is attributed to a causal relation between two
events, after coarse-graining. Of course, if
\begin{equation}
\sum_{i}^{}\left( V_{i} + K_{i} \right) = - \frac{\hbar\kappa\eta}{2\pi}
\end{equation}
where, the minus sign stands for showing the cohesive energy behavior of
the causal relation between the events of the denumerable causal
spacetime, then
\begin{equation}
\left( \frac{2\pi}{\hbar\kappa\eta} \right)T_{\text{ab}} = \{ R_{\text{ab}} + \left( - \frac{R}{2} + \Lambda \right)\} g_{\text{ab}}
\end{equation}
which is the Einstein state equation {[}12{]}. Using Eq. (26), it is
possible to obtain a limit that the discrete state (Eq. (23)) becomes a
continuous state (Eq. (27)).
\\
 \\
Since, \(k_{B}T\) has energy dimension, therefore
\(k_{B}(\frac{\hbar\kappa}{2\pi})\) has energy dimension. But,
\(\sum_{i}^{}\left( V_{i} + K_{i} \right)\) is energy per unit area
therefore the coefficient \(\eta/k_{B}\) has the inverse of the
dimension of area and
\(k_{B}\left( \frac{\hbar\kappa}{2\pi} \right)\left( \frac{\eta}{k_{B}} \right)\)
has energy per area dimension. Now, if
\(\frac{\eta}{k_{B}}\sim\frac{1}{l_{p}^{2}}\) where \(l_{p} =\)
\(\sqrt{G\hslash/c^{3}}\) is the Planck length then
\(\eta\sim\frac{k_{B}}{l_{p}^{2}}\). The result is in good agreement
with the Hawking's result which is \(\eta = \frac{k_{B}}{4l_{p}^{2}}\)
{[}36{]}.

Also, since the spacetime area $\delta A$ is composed by the causal
events, it may be possible one calculates the discrete Ricci curvature
in terms of the graph which is made by the causal events when the
background-independent coarse grain technique is applied {[}37, 38{]}
and then solve the matrix Eq. (24). It can be the subject of the future
researches.

\section{Summary}

\noindent

It was assumed that spacetime is composed by events (called parents).
The parents born children which some children have causal relation with
their parents and some have not. It was assumed that the only causal
children take part in the evolution of the system in each step of
growth. The system has been modeled by the Leslie matrix with finite
(infinite) dimensions. In both finite and infinite dimension, it has
been shown that the stationary population of causal events exist and for
the population the relation $\Delta S = - \Delta Q$ is satisfied where
\(S\) is the entropy of system and \(Q\) is its internal energy. For
preserving the casual relation between parents and children, some works
should be done which is stored in the system as internal energy
\(\Delta Q < 0\). Since, there are denumerable casual relations, by
assuming that the stored energy is equal to \(\hslash\omega/2\) for each
causal relation, it has been shown that the energy of vacuum is equal to
\(m\hslash\omega/2\) where \(m\) is the number of casual parents which
took part in the evolution of the system and in consequence the entropy
is quantized. Also, we assumed that \(\delta A\) area of spacetime is
composed by the disordered causal events which have the causal bounding
energy \(V_{i}\) and the kinetic energy \(K_{i}\). By considering the
past horizon of a local Rindler horizon which is instantaneously
stationary (in local equilibrium) at spacetime point \(p\), the relation
between energy-momentum tensor \(T_{\text{ab}}\) and Ricci tensor has
been found. The relation can be called the Einstein state equation.
Using the discrete curvature theory and the coarse grain technique, one
may solve the discrete state Einstein equation for specific shape of
denumerable causal spacetime (further future research subjects).
Finally, it was shown that the coefficient between entropy and area
i.e., \(\eta\sim\frac{k_{B}}{l_{p}^{2}}\) which is in good agreement
with the Hawking's result.
\\
 \\
\appendix
\section{}


For example, if
\(M=  \begin{bmatrix}
    m_{1} & m_{2} \\
    b_{1} & 0
  \end{bmatrix}
\)
it can be easily shown that, the eingenvalues are
\begin{equation}
E = \frac{m_{1}}{2} \pm \sqrt{\frac{m_{1}^{2}}{4} + m_{2}b_{1}}
: = \lambda_{1},\lambda_{2}
\end{equation}
and if \(\vec {u} = \begin{pmatrix}
u_{1} \\
u_{2} \\
\end{pmatrix}\) is eigenvector then
\begin{equation}
m_{1}u_{1} + m_{2}u_{2} = \lambda_{i}u_{1}
\end{equation}
\begin{equation}
b_{1}u_{1} = \lambda_{i}u_{2}
\end{equation}
 Therefore
\begin{equation}
\vec{u} = \begin{pmatrix}
u_{1} \\
u_{2} \\
\end{pmatrix} = \frac{1}{\sqrt{\lambda_{i}^{2} + b_{1}^{2}}}\begin{pmatrix}
\lambda_{i} \\
b_{1} \\
\end{pmatrix} = \frac{1}{\lambda_{i}^{2}}\frac{1}{\sqrt{\lambda_{i}^{2} + b_{1}^{2}}}\begin{pmatrix}
1/\lambda_{i} \\
b_{1}/\lambda_{i}^{2} \\
\end{pmatrix}
\end{equation}
and
\begin{widetext}
\begin{equation}
\begin{pmatrix}
m_{1} & m_{2} \\
b_{1} & 0 \\
\end{pmatrix}\begin{pmatrix}
1/\lambda_{i} \\
b_{1}/\lambda_{i}^{2} \\
\end{pmatrix} = \begin{pmatrix}
\left( m_{1}\lambda_{i} + b_{1}m_{2} \right)/\lambda_{i}^{2} \\
b_{1}/\lambda_{i} \\
\end{pmatrix} = \begin{pmatrix}
1 \\
b_{1}/\lambda_{i} \\
\end{pmatrix} = \lambda_{i}\begin{pmatrix}
1/\lambda_{i} \\
b_{1}/\lambda_{i}^{2} \\
\end{pmatrix}
\end{equation}
\end{widetext}
However, if \(\vec{v}M = \lambda\vec{v}\) and
\(\vec{v} = \begin{pmatrix}
v_{1} & v_{2} \\
\end{pmatrix}\) then
\begin{equation}
m_{1}v_{1} + b_{1}v_{2} = \lambda v_{1}
\end{equation}
\begin{equation}
m_{2}v_{1} = \lambda v_{2}
\end{equation}
Therefore
\begin{equation}
\vec{v} = \begin{pmatrix}
v_{1} & v_{2} \\
\end{pmatrix} = \frac{1}{\sqrt{\lambda_{i}^{2} + m_{2}^{2}}}\begin{pmatrix}
\lambda_{i} & m_{2} \\
\end{pmatrix}
\end{equation}
and
\begin{widetext}
\begin{equation}
\begin{pmatrix}
\lambda_{i} & m_{2} \\
\end{pmatrix}\begin{pmatrix}
m_{1} & m_{2} \\
b_{1} & 0 \\
\end{pmatrix} = \begin{pmatrix}
\lambda_{i}m_{1} + m_{2}b_{1} & \lambda_{i}m_{2} \\
\end{pmatrix} = \begin{pmatrix}
\lambda_{i}^{2} & \lambda_{i}m_{2} \\
\end{pmatrix} = \lambda_{i}\begin{pmatrix}
\lambda_{i} & m_{2} \\
\end{pmatrix}
\end{equation}
\end{widetext}
Now, if \(M^{*} = \begin{pmatrix}
0 & - m_{2} \\
 - b_{1} & m_{1} \\
\end{pmatrix}\) is the adjoint of \(M\) then
\begin{widetext}
\begin{equation}
\begin{pmatrix}
0 & - m_{2} \\
 - b_{1} & m_{1} \\
\end{pmatrix}\begin{pmatrix}
 - m_{2} \\
\lambda_{i} \\
\end{pmatrix} = \begin{pmatrix}
 - m_{2}\lambda_{i} \\
b_{1}m_{2} + m_{1}\lambda_{i} \\
\end{pmatrix} = \lambda_{i}\begin{pmatrix}
 - m_{2} \\
\lambda_{i} \\
\end{pmatrix}
\end{equation}
\end{widetext}
Therefore, \({\vec{v}}^{*} = \begin{pmatrix}
 - m_{2} \\
\lambda_{i} \\
\end{pmatrix}\) is the eigenvector of \(M^{*}\).
\\
 \\
\section{}
It has been shown that {[}34{]}
\\
 \\
\textbf{Theorem\#1:} Let \(M\) be an infinite-dimensional population
matrix. If (i) \(m_{i} > \ 0\) for infinitely many \(i\) and \(m_{i}\),
bounded, (ii) the greatest common divisor of the subscripts \(i\) of the
positive \(m_{i}\) is unity, and (iii)
\(\text{Limit}_{i \rightarrow \infty}b_{i} = 0\), then the population
matrix \(M\) has essentially unique stationary distribution
\(\overline{x} = (x_{1},\ x_{2},\cdots)\), \(x_{i} > 0\).
\\
 \\
And also {[}34{]}
\\
 \\
\textbf{Theorem\#2:} If (i) \(m_{i} > \ 0\) for infinitely many n, and
\(m_{i}\) bounded; (ii) the matrix \(M\) defines a compact operator
\(T\) on \(l_{2}\), that is the Hilbert space of vectors
\(\overline{x} = (x_{1},\ x_{2},\cdots)\) for which
\(\sum_{}^{}{\left| x_{i} \right|^{2} < \infty}\), then the equation
\(M\overline{x} = \lambda\overline{x}\) has a solution \(\lambda > 0\)
and \(\overline{x} = (x_{1},\ x_{2},\cdots)\), \(x_{i} > 0\).
\\
 \\
Therefore, for \(i \rightarrow \infty\) an essentially unique stationary
distribution exists {[}34{]}.

However, we can write the entries of an infinite denumerable Leslie
matrix as {[}35{]}
\begin{equation}
M_{\text{ij}} = \delta_{1j}m_{j} + \delta_{i,j + 1}b_{i}
\end{equation}
Therefore, the corresponding discrete dynamical equation is
\begin{equation}
u_{n} = M\ u_{n - 1}
\end{equation}
Now, if \(u_{0}\) is the first parent (initial condition) then
\begin{equation}
u_{n} = M^{n}\ u_{0}
\end{equation}
\textbf{Lemma:} It has been shown that, there is the generating
matrix \(G(z)\) such that {[}35{]}

\[G\left( z \right) = \sum_{n \geq 0}^{}{M^{n}\ z^{n}}\]

where, \(z\) is a complex indeterminate and the entries of \(G\) is
\begin{equation}
G_{\text{ij}} = C_{j}^{i - 1}z^{i - j} + \frac{C_{1}^{i - 1}\sum_{n \geq 0}^{}{m_{j + n - 1}C_{j}^{j + n - 1}z^{n + i}}}{\mathrm{\Delta}(z)}
\end{equation}
Here,
\begin{equation}
C_{k_{i}}^{k_{f}} = \left\{ \begin{matrix}
\Pi_{{k = k}_{i}}^{k_{f}}b_{k}\text{\ \ \ \ \ \ \ \ \ \ \ \ \ \ \ \ \ \ for\ }k_{f} > k_{i} \\
1\ \ \ \ \ \ \ \ \ \ \ \ \ \ \ \ \ for\ \ \ \ \ k_{f} = k_{i} - 1 \\
0\ \ \ \ \ \ \ \ \ \ \ \ \ \ \ \ \ \ \ \ \ \ \ \ \ \ \ \ \ \ \ \ \ \ \ \ \ Otherwise \\
\end{matrix} \right.\
\end{equation}
which is similar to Eq. (2) and
\begin{equation}
\mathrm{\Delta}\left( z \right) = 1 - \sum_{n \geq 0}^{}{m_{n}C_{1}^{n}z^{n + 1}}
\end{equation}
It should be noted that, the real root of
\(\mathrm{\Delta}\left( z \right)\ \)is associated with the solution of
Euler--Lotka equation which is {[}35{]}
\begin{equation}
1 - \sum_{n \geq 0}^{}{m_{n}C_{1}^{n}\left( \frac{1}{\rho} \right)^{n + 1} = 0}
\end{equation}
where, \(\rho\) is the leading eigenvalue (\(\rho > 0)\).

\section{}

Let us to consider a black hole event horizon. The system is the degrees
of freedom beyond the horizon and the outside world is separated from
the system by a causality barrier. Such a system is not in equilibrium
because the horizon is expanding, contracting, or shearing. A local
Rindler horizon of a small spacelike 2-surface element$P$ whose past
directed null normal congruence to one side (which we call the inside)
and has vanishing expansion and shear at a first order neighborhood of
each spacetime point$P$ is considered {[}12{]}. The part of spacetime
beyond the Rindler horizon that is instantaneously stationary (in local
equilibrium) at $P$ is considered as system. The heat flux to the past
is given by {[}12{]}
\begin{equation}
\delta Q = \int_{}^{}{T_{\text{ab}}\chi^{a}d\Sigma^{b}}
\end{equation}
where, \(T_{\text{ab}}\) is the matter energy-momentum tensor and
\(\chi^{a}\) is the killing field generating boost orthogonal to $P$
and vanishing at $P$. The integral is over a pencil of generators of
the inside past horizon of $P$. Of course, it is assumed that the
temperature of the system is equal to the Unruh temperature which is
observed by observer hovering just inside the horizon. It can be shown
that {[}12{]}
\begin{equation}
\delta Q = - \kappa\int_{}^{}{\lambda T_{\text{ab}}k^{a}k^{b}{d\lambda dA}}
\end{equation}
where, \(k^{a}\) is the tangent vector to the horizon generators for an
affine parameter \(\lambda\) that vanishes at $P$ and is negative to
the past of $P$, \(\chi^{a} = - \kappa\lambda k^{a}\) and
\(d\Sigma^{a} = k^{a}{d\lambda dA}\) , where $dA$ is the area
element on a cross section of the horizon. Here, \(\kappa\) is the
acceleration of the Killing orbit on which the norm of \(\chi^{a}\) is
unity and it is assumed that the speed of light is equal to unity. However, according to the Unruh effect, the Minkowski vacuum state of quantum fields is a thermal state with respect
to the boost Hamiltonian at temperature \(T = \hslash\kappa/2\pi\) and
it can be shown that {[}12{]}
\begin{equation}
\delta Q = \left( \frac{\hbar\kappa}{2\pi} \right)\eta\int_{}^{}\lambda\ R_{\text{ab}}k^{a}k^{b}{\lambda dA}
\end{equation}
\\
Therefore, for all null \(k^{a}\) {[}12{]}
\begin{equation}
\frac{2\pi}{\hbar\eta}T_{\text{ab}} = R_{\text{ab}} + ( - \frac{R}{2} + \Lambda)g_{\text{ab}}
\end{equation}
where, \(R_{\text{ab}}\) and \(R\) are Ricci tensor and Ricci scalar,
respectively. \(\Lambda\) is some constant. Here, it is assumed that
{[}12{]}
\begin{equation}
\delta A = - \int_{}^{}{\lambda R_{\text{ab}}k^{a}k^{b}{\lambda dA}}
\end{equation}
which is the area variation of a cross section of a pencil of generators
of the inside past horizon of $P$.

\nocite{*}

\bibliography{apssamp}
 \textbf{References}

\noindent [1]C. S. McCaw, ``Orbitals: With Applications in Atomic
Spectra'' (Imperial College Press,
2015).

\noindent [2] P.~M.~Marcus, J.~F.~Janak, and A.~R.~Williams,
``Computational Methods in Band Theory'' (Springer,
1971).

\noindent [3] J. C. Inkson, ``many-body theory of solids'' (Springer
US,
1984).

\noindent [4] Yuli V. Nazarov, ``Quantum Transport: Introduction to
Nanoscience'' (Cambridge University Press,
2009).

\noindent [5] Supriyo Datta, ``Quantum Transport:
Atom to Transistor'' (Cambridge University Press, 2005).

\noindent [6]W. Greiner, S. Schramm, and E. Stein, ``Quantum
Chromodynamics'' (Springer,
2007).

\noindent [7]E. A. Paschos, ``Electroweak Theory''
(Cambridge University Press, 2005). 

\noindent [8]  M. D. Schwartz, ``Quantum Field Theory and the Standard Model'' (Cambridge University Press, 2013).

\noindent [9] Sean M. Carroll, ``Spacetime and Geometry: An
Introduction to General Relativity'' (Addison-Wesley,
2003).

\noindent [10] John Stewart, ``Advanced General Relativity'' (Cambridge University Press, 2009). 

\noindent [11]  S. W. Hawking,  ``Black hole explosions?" Nature, \textbf{248}, (1974) 30.

\noindent [12] T. Jacobson, ``Thermodynamics of Spacetime: The
Einstein Equation of State'' Phys. Rev. Lett. \textbf{75}, (1995) 1260.

\noindent [13] Bernd Schroder, ``Ordered Sets'' (Springer,
2016).

\noindent [14] A. A. Robb,  ``A Theory of Time and Space'' (Cambridge
University Press, 1914).

\noindent [15] A. A. Robb,``Geometry of Time and Space'' (At the University
Press, 1936).

\noindent [16] S. W. Hawking, A. R. King, and P. J. Mccarthy, ``A New Topology
for Curved Space-Time which incorporates the causal, differential, and
conformal structures'', J. Math. Phys. \textbf{17}, (1976) 174.

\noindent [17] D. B. Malament, ``The class of continuous timelike curves
determines the topology of spacetime'', J. Mathem. Phys. \textbf{18}, (1977)
1399.

\noindent [18]L. Bombelli, J. Lee, D. Meyer, and R. Sorkin, ``Space-Time as a
Causal Set'', Phys. Rev. Lett. \textbf{59}, (1987) 521.

\noindent [19] G. \textsuperscript{,}t Hooft, ``Quantum Gravity: A Fundamental
Problem and Some Radical Ideas'', Springer US, Boston \textbf{323} (1979).

\noindent  [20] J. Myrheim, ``Statistical Geometry'', CERN-TH-2538.

\noindent [21] S. Surya, ``The causal set approach to quantum
gravity'' Living Review in Relativity \textbf{22}, (2019) 5.

\noindent [22] R. D. Sorkin, ``Causal sets: Discrete gravity'',
Lecture on quantum gravity, Valdivia, Chile
(2002).

\noindent  [23] D. P. Rideout and R. D. Sorkin, ``A Classical
sequential growth dynamics for causal sets'', Phys. Rev. D \textbf{61}, (2000) 024002.

\noindent [24] Benjamin F. Dribus, ``Discrete Causal Theory''
(Springer,
2017).

\noindent [25] M. Cortes and L. Smolin, ``The Universe as a process
of unique events'' Phys. Rev. D \textbf{90}, (2014) 084007.

\noindent [26] M. Cortes and L. Smolin, ``Quantum Energetic Causal
Sets'' Phys. Rev. D \textbf{90}, (2014) 044035.

\noindent [27] Hamidreza Simchi, ``The Concept of Time: Causality,
Precedence, and Spacetime'', arXiv:2108.01353v1
(2021).

\noindent [28] Paul~Cull, Mary~Flahive, and Robby~Robson,
``Difference Equations'' (Springer,
2005).

\noindent [29] Michel Rigo, ``Advanced Graph Theory and
Combinatorics'' (Wiley Online Library,
2016).

\noindent [30] Lloyd Demetrius ,'' Natural Selection and Age-Structured Populations'', Genetic \textbf{79}, (1975) 535 .

\noindent [31] J. S. Dugdale, ``Entropy and Its Physical Meaning''
(Taylor \& Francis,
1996).

\noindent [32] Thomas M. Cover and Joy A. Thomas, ``Elements of
Information Theory'' (John Wiley and Sons
2006).

\noindent [33] S. Mani Naidu, ``Applied Physics'' (Dorling
Kindersley,
2010).

\noindent [34] Lloyd Demetrius, ``On an Infinite Population Matrix'',
Mathematical Bioscience \textbf{13}, (1972) 133.

\noindent [35] Joa˜o F. Alvesa, Anto´nio Bravob, and Henrique M.
Oliveira, ``Population dynamics with infinite Leslie matrices: finite
time properties'', J. Differ. Equat. and Appl. \textbf{29}, (2014) 9.

\noindent [36] S. Carlip, ``Black Hole Thermodynamics'', International Journal of Modern Physics D \textbf{23}, (2014) 1430023.

\noindent [37] Jean-Michel Morel and Bernard Teissier, ``Modern Approaches to Discrete Curvature'' (Springer, 2017).

\noindent [38]D. P. Rideout and R. D. Sorkin,'' Evidence for a continuum limit in causal set dynamics'', Phys. Rev. D \textbf{63}, (2001) 104011.

\noindent 

\noindent

\end{document}